# SERS study of single-live-cell electrical permeabilization dynamics via plasmonic nanotubes

Yuge Liang[1,3], Peilin Xin[1,3], Enock Adjei Agyekum[4], Jiaming Zhang[5,6], Yingqi Zhao[1,3], Jinglai Duan[5,6], Francesco De Angelis[7], Aki Manninen[2,3], Jianan Huang[1,2,3]*

[1]Research Unit of Health Sciences and Technology, Faculty of Medicine, University of Oulu, Aapistie 5 A, 90220 Oulu, Finland.

[2]Research Unit of Disease Networks, Faculty of Biochemistry and Molecular Medicine, University of Oulu, Aapistie 5 A, 90220 Oulu, Finland.

[3]Biocenter Oulu, University of Oulu, Aapistie 5 A, 90220 Oulu, Finland.

[4]The Biomimetics and Intelligent Systems (BISG) research unit, Faculty of Information Technology and Electronic Engineering, University of Oulu, Erkki Koiso-Kanttilan katu 3, 90570 Oulu, Finland.

[5]State Key Laboratory of Heavy Ion Science and Technology, Institute of Modern Physics Chinese Academy of Sciences, Lanzhou 730000, China.

[6]Advanced Energy Science and Technology Guangdong Laboratory, Huizhou 516000, China.

[7]Istituto Italiano di Tecnologia, Via Morego 30, 16163 Genoa, Italy.

*Email: jianan.huang@oulu.fi

Abstract: There is a growing demand for minimally invasive methods to analyze intracellular processes and signaling activities in individual living cells, including the identification of tumorigenic cell subpopulations. However, most conventional analytical methods require cell lysis, precluding repeated measurements in the same cell over time, or rely on exogenous labels and reporters that may perturb cellular function. Various applications based on vertical nanotubes have been developed that enable live cell monitoring and analysis by electroporation with low voltages. However, the extent and duration of membrane permeability and kinetics of membrane repair remain elusive. Here, we built a plasmonic platform with the capacity of surface enhanced Raman spectroscopy (SERS) to monitor the electroporation-induced membrane permeability dynamics in individual live cells attached onto 100-nm diameter nanotubes of 2 μm height. Fibronectin was employed as extracellular matrix (ECM)-coating to facilitate cell attachment onto nanotubes. Using fluorescent-dye delivery as an independent validation method, we show that the fabricated nanostructures induce localized electrical permeabilization of the plasma membrane and enable monitoring of its subsequent recovery. We further use SERS to track molecular changes at the membrane during permeabilization and resealing. The SERS spectra provide molecular-level insight into changes in membrane-associated components and the ECM during electroporation and subsequent membrane recovery. Real time monitoring of pulse induced molecular changes holds great promise for characterizing intracellular signaling, cellular states, and cellular heterogeneity at the single cell level, including the identification of tumorigenic subpopulations. This capacity could facilitate the development of novel biosensing assay.

## Introduction

Single-cell analysis has been advanced as a crucial technique that allows researchers to investigate the study of genomics, transcriptomics, proteomics, metabolomics and extracellular interactions at an individual cell level, which brings a new insight into heterogeneity, cellular mechanisms and treatment responses in cancer study[1–3]. The extracellular matrix (ECM) regulates tumor cell signaling and invasiveness. The ECM molecules are bound by integrins that are cellular ECM receptors serving as a bidirectional link between the cells and the ECM to regulate cellular functions such as migration, differentiation, survival, and tissue homeostasis[4–6]. The plasma membrane is an essential regulatory barrier that maintains cellular homeostasis but also poses challenges for real-time monitoring of intracellular processes and the delivery of therapeutic molecules into cells[7,8]. Conventional cytoplasmic sampling techniques usually rely on destructive methods that lead to cell lysis excluding chronological analysis at single-cell level[9,10]. Nanostructure platforms provide an effective and minimally invasive monitoring approach to tackle this issue. Present related cellular real-time sensors have been classified as ‘on cell’ in terms of spatial contact[11]. Various techniques have been proposed as ‘on-cell’ sensors direct contacting the cell membrane, thereby balancing between minimal invasiveness and high-resolution detection implemented through transient permeabilization of the plasma membrane using techniques such as mechanoporation and electroporation[12–18]. Electroporation, or electropermeabilization, is a commonly used strategy to increase cell membrane permeability by transiently or permanently impairing membrane integrity through electrical fields. Nanoscale electroporation techniques present a promising approach with significant advantages over traditional bulk electroporation methods due to gentle electric fields applied to target cells where plasma membrane is only locally perturbed[19,20]. Spatially restricted electrical permeabilization using nanostructures makes it possible to intracellular probing from live cells[12,21–27]. Fluorescence-based sensing is widely used because of its high specificity, real-time imaging capability, and methodological maturity. However, fluorescence-based methods generally require fluorescent labels or reporters, and prolonged or repeated imaging may cause photobleaching and phototoxicity, potentially compromising cell viability and limiting longitudinal analysis. Electrophysiological recordings enable sensitive monitoring of the surrounding microenvironmental changes, but such analyses suffer from limited spatial resolution and throughput. Raman spectroscopy reflects the molecule’s vibrational modes and surface-enhanced Raman spectroscopy (SERS) offers high resolution, spatial information, ultrahigh sensitivity, and hence enables universal and rapid molecular identification, eliminating the need to label specific biomarkers and showing promise in live-cell analysis such as detection of extracellular potassium concentrations using ionophore-containing polyvinyl chloride films[28]and analyzing membrane-associated biomolecules via plasmonic nanopillars[29]or nanoparticles[30].

Several research utilizes SERS employing plasmonic nanostructures to acquire real-time molecular analysis inside the cell, for example via intracellular nanorods[31], SERS analysis of intracellular compounds[32] and long-term monitoring of molecular pathological processes on plasmonic microchips[33]. Existing SERS platforms are restricted by shallow depth and strong reliance on near-field enhancement from thin-film; and pulse-induced cell morphology changes and detachment have been observed on SEM images, which limit SERS optimal performance[34]. Present membrane-focused SERS approaches often use functionalized tags to detect specific molecules. However, adherent cells require ECM-mediated attachment to interact with artificial surfaces, the resulting interfacial layer can interfere with direct membrane-substrate contact and complicate label-free analysis of membrane-associated molecules. Fibronectin (FN) is a major component of ECM which binds to integrin $\alpha_v\beta_3$ and integrin $\alpha_5\beta_1$ from cell membrane via its RGD (Arg-Gly-Asp) motif[35–40]. Integrin-mediated interactions are required for active adhesion, and therefore research groups employed RGD peptides to promote cell-substrate interactions. Furthermore, cell membrane electropermeabilization mechanisms is still unknowing, even though lipid bilayers aqueous pores generation have been proved[41–43]. Membrane protein conformational changes may serve as available biosignatures, which is supported by previous voltage-gated ion channels investigations that indicated electroconformational changes possibly contribute to the permeabilized state[44–47]. These observations raised the question of how ECM influences electroporation-induced molecular changes in the plasma membrane and its subsequent recovery. It is also imperative to understand the molecular mechanisms of membrane permeabilization and resealing under physiologically relevant conditions. SERS-active nanostructures offer a potentially powerful platform for monitoring these processes in real time with high sensitivity, without requiring molecular labels and with minimal perturbation of the cell[48,49].

Herein, we developed a multifunctional 'on cell' plasmonic nanotube-array sensor to monitor the real-time dynamics of electropermeabilized plasma membranes with SERS. As shown in Figure 1, nanotubes are grown on a designed-etched $Si_3N_4$ membrane, a circular pore leads to nanoscale gaps on the structure tip that generates hotspots, the plasmons can be propagated along the tube surface on which PC-3 cancer cells have adhered allow Raman signals of the proximal cell membrane to be enhanced. We employed FN proteins to facilitate robust cell adhesion during electropermeabilization and subsequent analysis. The fabricated hollow metal-sputtered nanostructures can be treated as nanoscale electrodes to spatially electroporate the cell membrane. These features promote visualization of cell permeabilization through delivery of fluorescent dyes into calcein-AM contained cells at defined time points, which helps to calibrate the permeabilization extent between SERS studies which concluded that cell membrane was spontaneously permeabilized and repaired. According to cell membrane electroporation dynamics, this time-resolved permeabilization also permits more complex patterns, that might help to adjust delivery of different amounts of cargo into cell via our adhesion-promoting nanoplatform. This technique has potential to get insight into how the ECM responds to electroporation and thereby perform detailed studies to explore cell permeabilization profiles that could be beneficial for accelerating drug delivery.

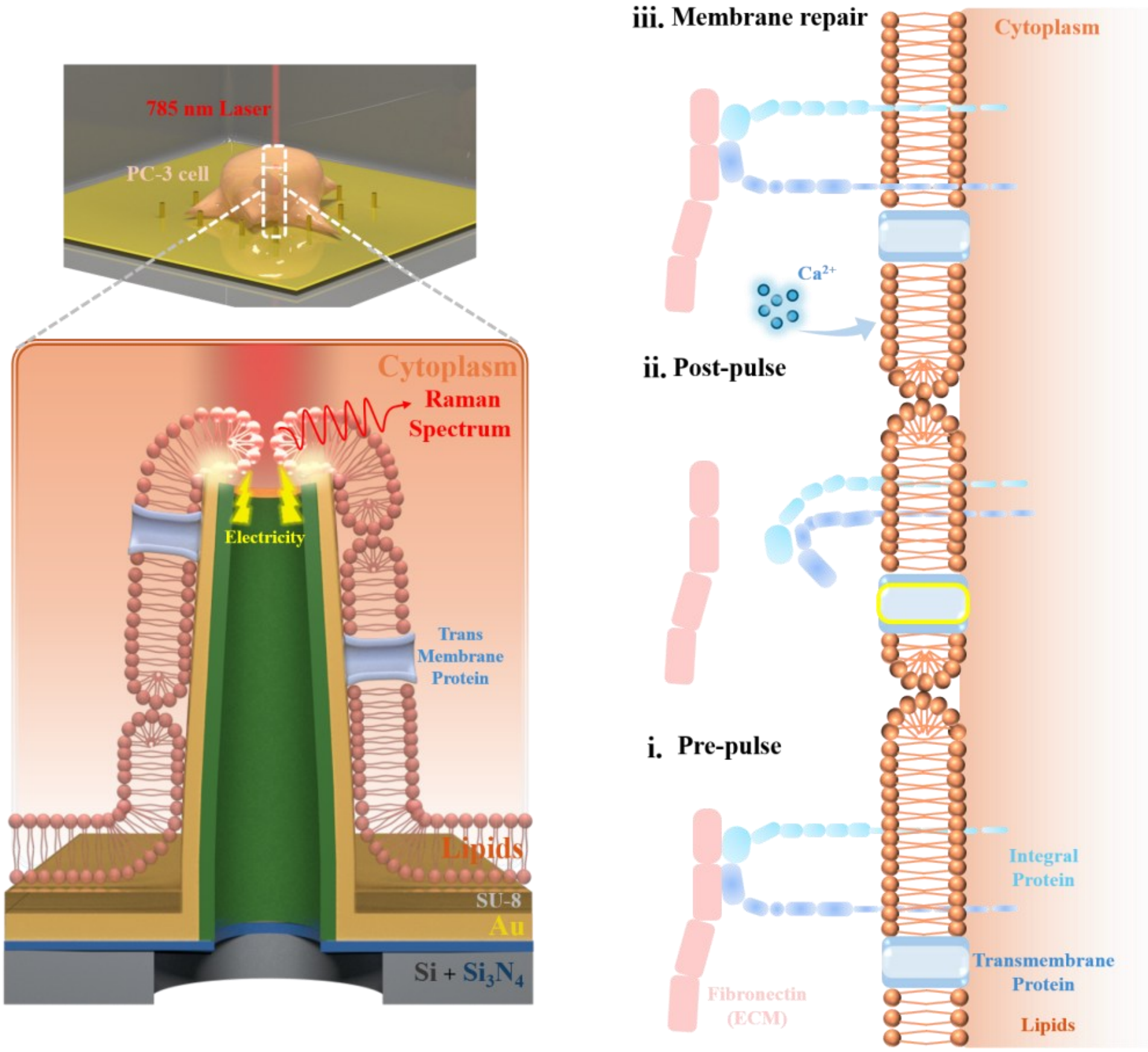

Figure 1. Schematic of nanotubes array fabrication with FIB milling, and electrical permeabilized ECM-single cell integrin engulfed on plasmonic nanotube.

## Results and discussion

### *Nanostructure parameters correlate to plasmonic effect*

Nanotubes, as a type of hollow structure, the nanotips combine enhanced Raman signals generation and electrical pulse applied nanoelectrodes; besides, various molecules are allowed to be delivered into cellular interior through the inner channel. This multifunctional nanoplatform fabrication is based on photoresist lithography and focused ion beam milling (FIB) (see Figure S1).

As the electromagnetic field around plasmonic materials is not uniformly distributed but highly localized in spatially narrow regions (SERS hotspots), such as nanotips[50], the specific parameters of nanotubes which determine plasmon aggregation are essential as a key factor. Finite-difference-time-domain (FDTD) method was applied to simulate the near-field electric fields amplitude distribution in three-dimensional structures. Based on simulation results, the targeted structures correspond to Raman laser wavelength will be fabricated by modifying FIB parameters that surface plasmon resonance generated and intensive electric fields created, thus detecting Raman signals enhanced for spectral analysis.

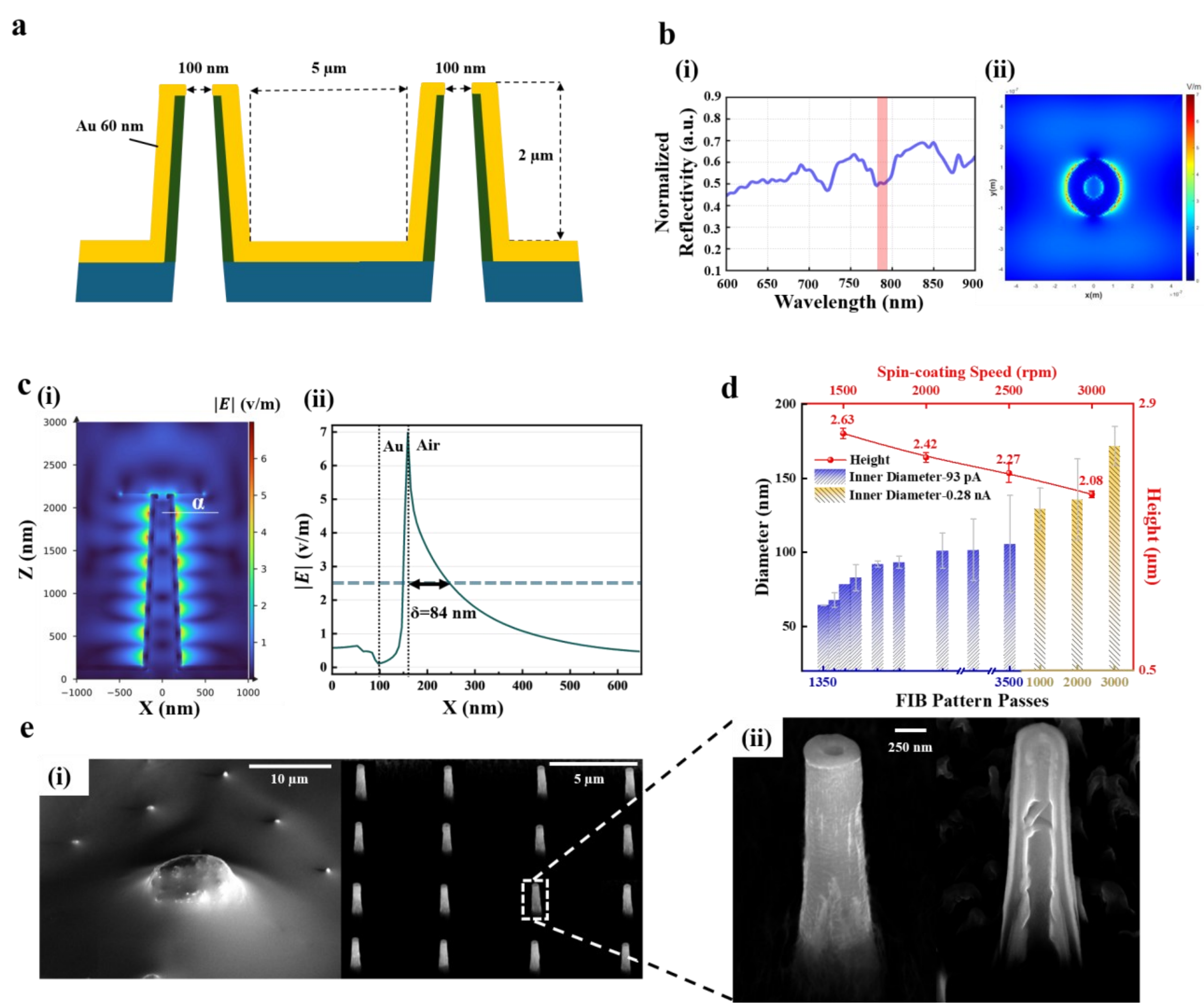

Figure 2. FDTD simulated nanotube array structure: a). schematic, b). Normalized reflectivity spectrum from 600 nm to 900 nm, red area locates illuminated Raman laser (i), electromagnetic intensities distribution at 785 nm of incident linearly polarized plane wave in top view (ii). c). FDTD simulated electric field intensity of nanotube at 785 nm in x-z plane (i), electric field intensity at position α, δ is the decay length of the surface plasmon which is defined as the intensity drops to its 1/e times of maximum value (ii). d). parameters collection of nanotubes based on FIB settings. e). FIB-SEM images of a single-cell gulfed on nanostructure surface and an exposed array (i), the magnificated images show a single tube and its cross-section (ii) (scale bar showed on images).

**Figure 2**a displays the designed structure array: 2 µm height, 100 nm inner diameter, 60 nm Au layer and 5 µm periodicities. Figure 2b-ii shows electromagnetic intensity distribution on nanotube under 785 nm laser illumination; the simulation exhibits the plasmonic field is reinforced and accumulated in a total volume that generated on the structure tip including inner circular hole surrounding, aggregated and propagated via the whole surface. As surface plasmon resonance generated, the near field was enhanced and surface plasmons was allowed to be excited corresponding to the minima in reflectivity spectrum (see Figure 2b-i, there is a red line marking 785 nm and an observed minimum in the spectrum, which agrees well with the validity of the FDTD simulation). As such the plasmon resonance position corresponded to incident laser wavelength was confirmed and near-electric field enhancement follows $|E(\omega)|^2|E(\omega')|^2/|E_0|^4$ ($E(\omega)$ and $E(\omega')$ represents the enhanced electric field amplitude at frequency $\omega$ and Raman emitted frequency $\omega'$ and $E_0$ stands for is the incident laser field amplitude) that the SERS enhancement factor is approximately proportional to the fourth power of the local electric field[51]. In Figure 2c, the enhanced field covers an area of ~2 µm in length on the structure with a decay length of 84 nm on the nanotip, which enables molecule-level spectral information acquired around this area. The wide-range fields are associated with propagating surface plasmons resonances of the continuous gold film. The Figure 2d showing collected parameters based on modifying fabricated protocol, increasing FIB ionic current and passes enlarges tube's inner diameter as well as higher spin-coating speed decreases photoresist thickness that controls tube's height based on collected parameters from **Figure S3**. As a result, 0.28 nA, 3000 passes and 3000 rpm can satisfy our demand to achieve specific-parameter structure.

The simulated results provide possibility to fabricate plasmonic nanoplatform as following collected parameters, which we use to demonstrate the availability of Raman signals enhancement. According to fabrication protocol, the morphology of structure has been exhibited via SEM imaging (**Figure S5**); the 4×4 nanotube arrays were fabricated and those parameters measurement correspond to theoretical results, which help resolve the discrepancy between theorical simulation studies and experimentally observed electron microscopy imaging. Since the pemeabilization must be achieved by electrical pulses, the SU-8 photoresist has been applied for passivation layer (see details on Methods) to insulate the gold planar surface as well as allows nanotubes exposed. Therefore, those structures can be treated as nanoelectrodes where the electric charge concentrates, thus generating a localized electrical field with a low amplitude voltage supplement. Figure 2e-i shows the SEM images that cell cultured on insulated nanostructure surface. Besides there is a hollow channel observed by using FIB milling, suggesting external molecules are allowed to be delivered into cells (see in Figure 2e-ii).

***Qualification and availability of plasmonic nanotubes array***

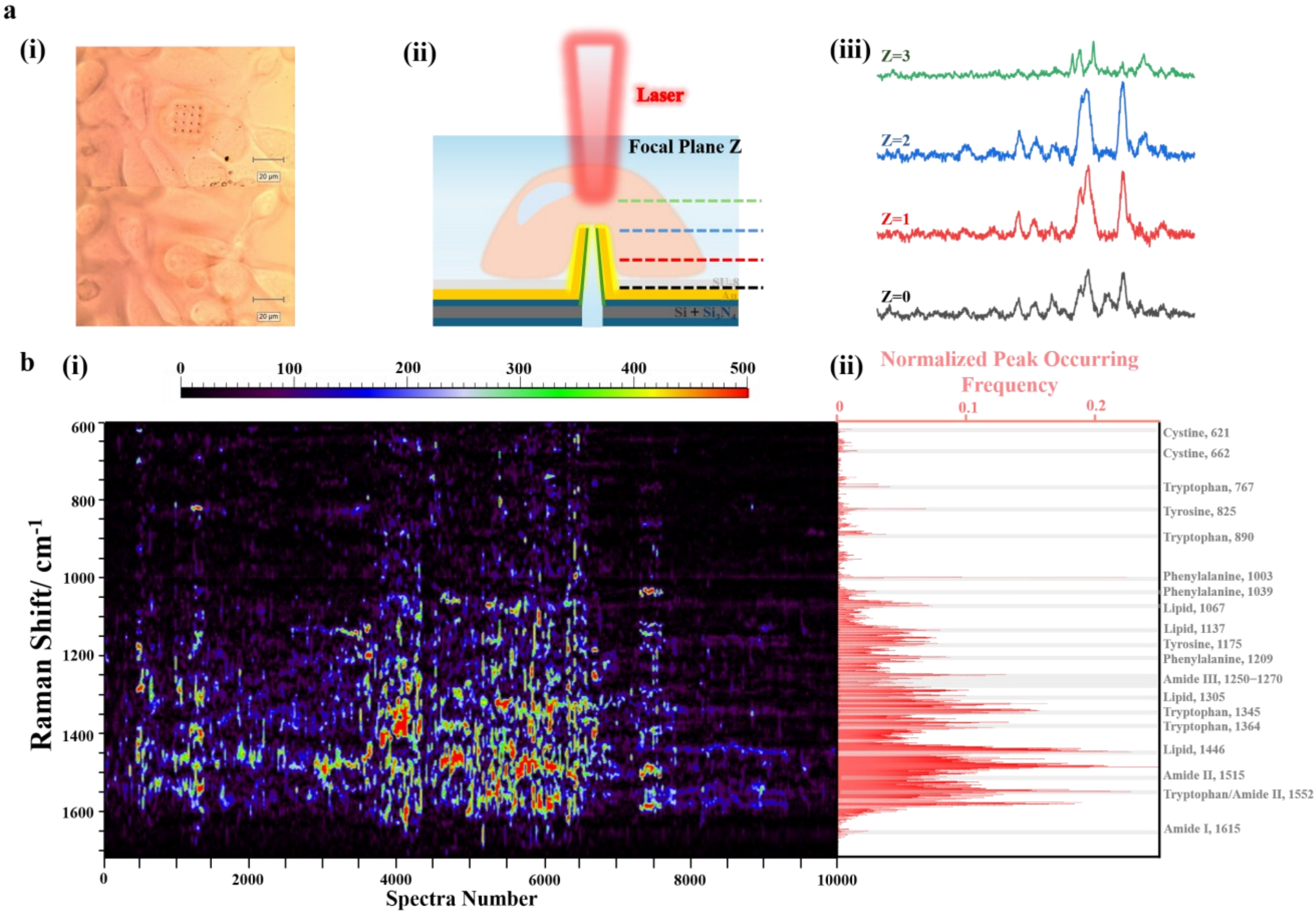

Figure 3. a) Microscope bright-field images of plasmonic nanotubes array and a single cell adhere to it (i). Schematic of different focal planes under laser illumination (ii). Collected spectra via different focal planes (iii). b) The 10000 Raman spectra are collected on nanotube arrays without ECM treatment after a single cell engulfed under the exposure time of 0.5 s, 9.5 mW of 785 nm laser (i). Distribution histograms of normalized peak occurring frequency of single PC-3 cell, and a single spectrum extracted from the data sets (ii).

Since plasmonic nanostructure has been made, its signal enhancement could be proved by depositing 4-Aminothiophenol (4-ABT) molecules that thiol group could be bonded to gold layer and generate SERS signals if plasmonic effect exists. And 4-ABT Raman signals were detected suggesting that fabricated nanotubes enable SERS capacity (see **Figure S6**). Thus, the SERS performance of plasmonic nanotube was evaluated in the absence of PC-3 cells. There are two images from Raman microscopy that demonstrate the appearance of nanostructure array and a cell seeded on it by changing the focus (**Figure 3**a-i). Since the plasmonic effect localized on the top from the three-dimensional feature, we optimized laser illumination by modifying focal plane, according to collected spectra from each focal plane and then determine the location (Figure 3a-ii). And the spectra corresponding each **Z** value has been shown in Figure 3a-iii, we selected the specific height (Z =2 μm) which performs optimal Raman intensity.

Next, SERS spectra of single PC-3 cell were recorded as a waterfall image in **Figure 3b**-i. The spectral profile, which is exclusively indicative of cell membrane constituents owing to the surface-selective properties of near-field plasmonic enhancement, exhibits common spectral membrane features as well as variable changes in real-time. The single living cell adhered poorly to non-ECM-coated nanostructures and resulted in variable SERS peaks, potentially reflecting intermittent exposure of different membrane-associated molecules to SERS hot spots. The weak adhesion led to fluctuations in peak position and intensity that invalidate distinguishing molecular identification, peaks occurrence frequency was utilized to statically determine the general spectral characteristics by analyzing 10000 SERS spectra in Figure 3b-i, as shown in Figure 3b-ii. Collected raw data were processed cosmic-ray-removal, subtract baseline and normalization in advance. Those features can be attributed to the different vibrational modes of proteins and lipids molecules. The most prominent bands appeared at 662 $cm^{-1}$ (C−S stretching in cystine)[52], 1003 $cm^{-1}$ (benzene ring breathing of phenylalanine)[53], the bands at 1067, 1137, 1305, and 1446 $cm^{-1}$ are mainly attributed to lipid hydrocarbon chain vibrations[54]. The peaks in the regions of 1250–1270, 1515–1552, and 1615 $cm^{-1}$ are related to protein amide III, amide II, and amide I vibrations, respectively[55]. Peaks assignments of all observed SERS bands are summarized in **Table S1**. This result suggests that by attaching single nanotube to the membrane, it is possible to obtain spectral information from this essential and complex structure.

***Kinetics of cell membrane electrical permeabilization and restoration***

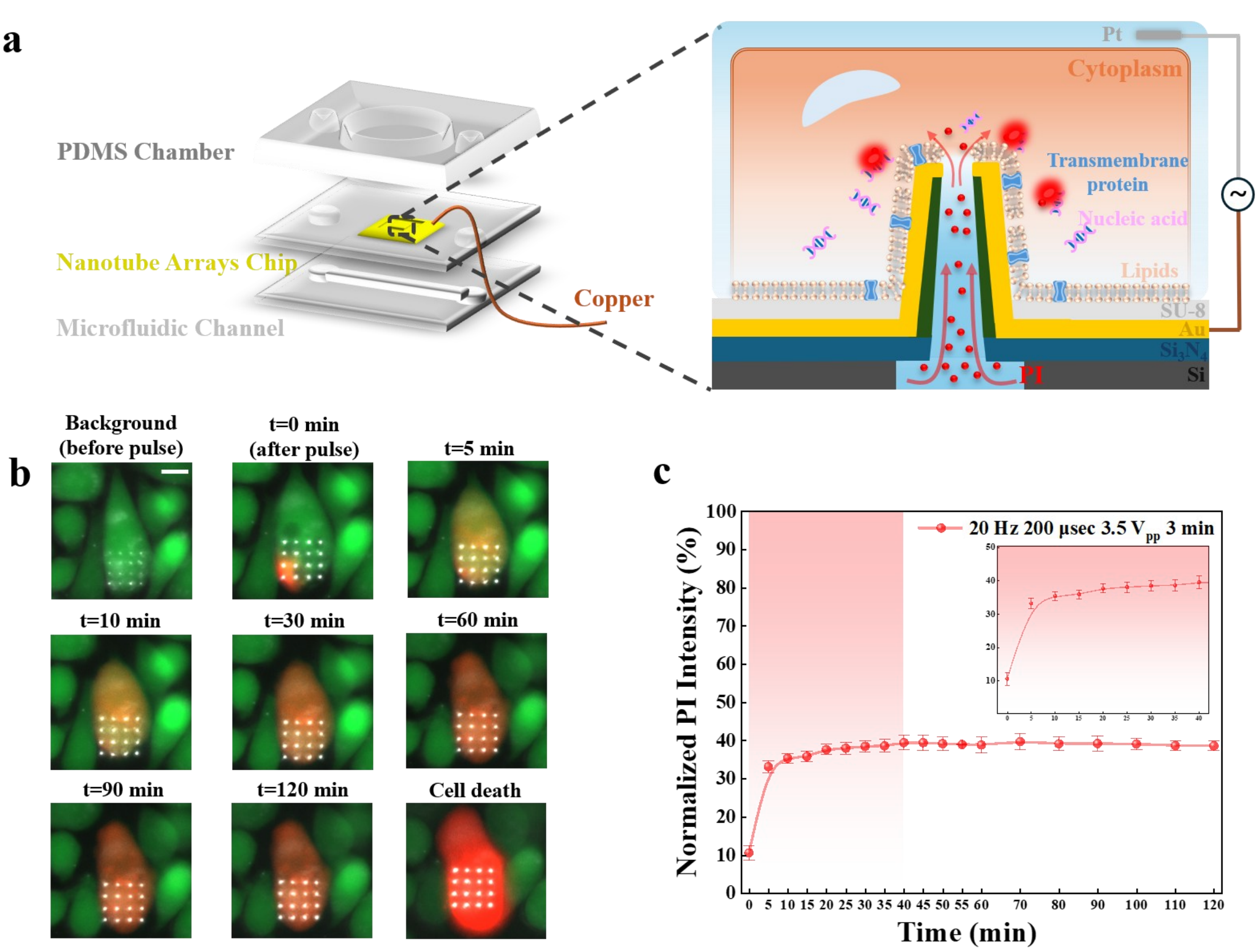

Figure 4. a). Schematic of fabricated PDMS chamber and PI delivery. b). Fluorescent images of calcein-AM expressing PC-3 cells seeded in culture wells following delivery of PI (scale bar represents10 μm). c). Normalized PI intensity after pulse supply following time (the time when pulse supply stopped defines as zero).

FN was used to functionalize the plasmonic nanotubes and promote cell adhesion, thereby improving contact between the analyzed cell and the SERS active substrate. Successful pulse-induced membrane permeabilization and intracellular delivery were assessed using propidium iodide (PI). The appearance of intracellular PI fluorescence indicated that the dye entered the cell through the nanotube-based delivery interface following electrical stimulus-induced local membrane permeabilization. **Figure 4a** shows the nanotubes membrane chip with PDMS chamber designed for PI dyes delivery; the membrane chip was attached with a copper cable as electrode and embedded in PDMS chamber with a delivery channel running below it. SU-8 passivation layer was deposited on gold surface allowing exposed nanotube conducted and rest flat area insulated that reinforce the localized electrical field during pulse applied. The designed device demonstrates the sputtered Au layer can be treated as electrode, the whole electrical field will be applied between the nanostructure and extracellular matrix as Pt wire was another electrode, which allows PI molecules to diffuse into permeabilized cells down their concentration gradient. After applying electrical pulse (20 Hz, 3.5 $V_{pp}$, 200 μsec) for 3 min, fluorescent images were collected in **Figure 4b**. The nanotube arrays could be observed on bright field channel and a single cell had been contained with calcein-AM that was observed as green pattern under fluorescence microscope and adhered on fabricated array. As electrical pulse supplement stopped, the image when the time was defined as zero shows intracellular PI signal upon binding with nucleic acids, indicating membrane permeabilization. The observed PI staining initiated from single nanotube and spread into the entire cell suggesting spatially restricted permeabilization and eventual spread and increase of PI staining intensity. **Figure 4c** exhibits the normalized PI intensity changes after pulse according to calculation (more details on *Methods*), the intensity increased until the PI level kept stable after 40 minutes, suggesting cell membrane electroporated and kept permeable allowing fluorescent dyes diffusion after pulse vanished until membrane completely resealed. Fluorescent interaction produces a pattern of spots in the cells, each one of which corresponds to a local membrane permeabilization event as PI is not membrane-permeable and diffused into cells only upon electroporation via the nanotube openings. Calcein-AM staining in cells indicated that nanoelectroporation did not significantly affect cell viability despite the observed PI delivery.

Since PI fluorescent dyes provide a rapid in situ permeabilization assay, PI intensity can be used to study permeabilization kinetics. By controlling molecular delivery into living cells upon pulse-induced permeabilization, the time frame for cell membrane permeabilization was ascertained as cells settled onto the nano structural substrate and observed in real time when PI was delivered. With strong pulse supplement, successful permeabilization became more apparent with intense PI signal.

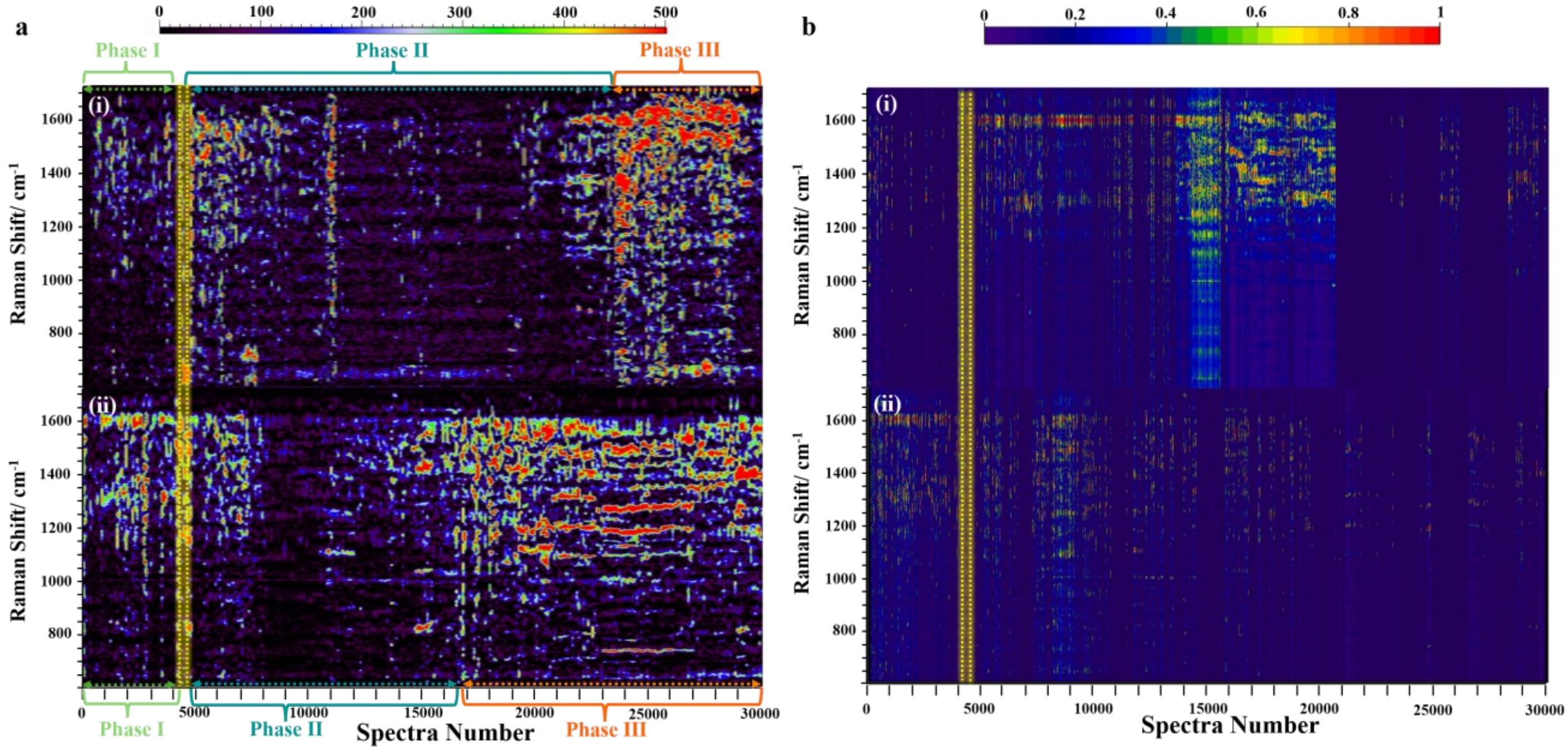

Figure 5. a). The SERS spectra of electrical permeabilized FN-cell system, which PC-3 cells were cultured in negative PBS (i) and positive PBS solution (ii). Yellow dot lines are marked as pulse lasting 300 spectra and we defined three phases in terms of electrical pulse supplement and SERS intensity. b). The SERS spectra of CNN-filtered data classification on PC-3 cell membrane, where collected in negative PBS (i) and positive PBS solution (ii).

The real-time SERS spectral monitoring with plasmonic nanotubes provides membrane dynamic changes over time as compared with other platforms, which we apply to demonstrate dynamics of cell membrane permeabilization. We employed FN as an ECM substrate facilitating cell adhesion to nanostructure surfaces to minimize cell detachmented from plasmon-resonance enhanced field. In terms of non-homogeneous distribution of analyte molecules and constrained near-field enhancement in nanotube, the thickness of biomolecular clusters including FN and its cellular integrin receptor and the plasma membrane is approximately 35−40 nm[56,57], suggesting that the decay length of electric field in plasmonic nanotube is broad enough to cover all the membrane proximal biomolecules.

To validate SERS monitoring contains membrane permeabilization and restoration, we incubated cells with PBS solution with ($PBS^+$) or without ($PBS^-$) $Ca^{2+}$ that is critical ion supporting plasma membrane repair. Calcium influx through membrane wounds is required for the repair process including localized exocytosis of lysosomes and lesion endocytosis[58]. The SERS spectra of electrical permeabilized FN-cell systems, as shown in Figure 5a, exhibit spectral-molecule dynamics as pulse is supplied. In **Figure 5**a-i, significant differences were observed between the spectra acquired pre-pulse and post-pulse measurements that we define three phases respectively regarding time trace and Raman intensity. As we applied pulse lasting 300 spectra marked as yellow dot line that determines Phase I before pulse application. After electro permeabilization applied, the intensities keep stable for around 2000 spectra, then there is dramatically decline in signals after electroporation applied, showing much lower intensities than before (Phase II), which mainly due to the sudden change in the environment. This drastic difference in the spectra lasted for 16000 spectra, after which the collected spectra began to increase showing more complex pattern (Phase III). We hypothesize that this is the process of membrane permeabilization and repair that FN-integrin binging detected at initial state; after pulse supplied, the connection was disrupted and recombined until membrane repaired. To validate the assumption, we changed to $PBS^+$ buffer and observed efficient membrane resealing as Figure 5a-ii shows. It is clear that the dynamics are similar in positive $PBS^+$ environment but Phase III can be observed more earlier and lasting longer, suggesting that $Ca^{2+}$ engages the Phase III which is determined by membrane repair.

Since the setup includes ECM-cell interface, it is clear that FN-coating and cellular integrin receptors binding to FN will interfere with the analysis of other membrane constituents. Hence, we applied a CNN model to predict and filter out membrane SERS spectra in Figure 5a trying to detect cellular components dynamics under electric fields. The SERS spectra of FN molecules have been collected as shown in Figure S7. The cell membrane SERS data in Figure 3b has been acquired as training feature to represent stable-state SERS signals and analyze the electrical permeabilized SERS data. Figure 5b shows the stable-state membrane signals distribution, the stable-state membrane features are observed during pre and post pulse, but there is no signature on Phare III due to other components contribute to high-intensity spectra that we hypothesis it belongs to repaired integrin-fibronectin interaction.

## Conclusion

In this work, we used Raman spectroscopic analysis of individual live cells to determine electrical plasma membrane permeabilization dynamics using plasmonic nanotube array platform with SERS capacity. The 2-μm height, 100-nm diameter nanotube array was fabricated by precisely controlling spin-coat speed, ionic current and passes on FIB milling following FDTD simulation that surface plasmon resonance can be generated under specific Raman laser enabling cell membrane signals collected within a highly confined plasmonic enhanced electromagnetic field. The cell membrane fluctuations make SERS spectra peaks and intensity various randomly; thus, the peak occurrence frequency was applied to detect biological components statically in label-free manner. Lipid molecules, peptides and protein backbone which belong to cell membrane were assigned and distinguished. PI dye was delivered via nanotubes into permeabilized cells to validate that the platform could be used to electrically permeabilize cells to allow intracellular delivery of PI. The normalized intracellular PI fluorescence increased by approximately 40% after pulse application and reached a plateau after 40 min. It should be noted, however, that this plateau is only a rough estimate as it cannot by itself be interpreted as evidence of plasma-membrane recovery, because PI fluorescence may stabilize following saturation of accessible intracellular nucleic-acid binding sites. Further methods are needed to pinpoint when the membrane repair process is complete. We thus monitored this process via SERS spectra in real-time. There are obvious drastic Raman intensities changes during the electrical permeabilization, and this process has been proved as membrane resealing by engaging $Ca^{2+}$. These observations promote determination of the function of ECM during electrical permeabilized membranes and get insight into membrane molecules dynamics which helps understand membrane repair mechanisms. The presented SERS technique could be a promising approach to investigate tumor cells behavior under multiple conditions and high potential to cell profiling that can be applied in disease diagnosis and precise treatment.

## Methods

*Materials:* silicon nitride wafer: Prime Si+$Si_3N_4$ wafer 4-inch, thickness 525±15 µm (100), 2-side polished, p-type (Boron), TTV<5 µm, 1-10 Ohm cm, 100 nm low-stress LPCVD $Si_3N_4$ on both sides, Flats: 2 (MicroChemicals). Titanium sputter target (70-TI5710, Micro to Nano BV), gold sputter target (70-AU5708, Micro to Nano BV).

*Device fabrication:* Photoresist S1813 (Microposit G2, Dupont) was spin-coated on a $Si_3N_4$ membrane at 3000 rpm for 1 min and soft baked at 110 °C for 1 min. After sputtering 2 nm thick titanium and 50 nm thick gold layer on the backside of the $Si_3N_4$ membrane, focused ion beam milling (FIB, Helios Nanolab 600 Dualbeam, FEI) was applied to fabricate arrays. Structure-formed chips were coated with a 2 nm titanium and 60 nm gold layer, then annealed at 200 °C for 1 hour (see details on Supporting Information). Photoresist SU-8 (TF 6001, Kayaku AM) was coated on the structural chip at 6000 rpm for 1 min, etched down by oxygen plasma at 200 W for 5 min to form insult layer and let nanotubes exposed, then hard baked at 200 °C for 2 hours. The microfluidic chamber was made in polydimethylsiloxane (PDMS), and a glass ring was bonded as a container for separating cells from outside. A copper wire was attached to the gold layer of chips with silver paste for connecting pulse generator and dropping Pt wire as negative electrode during pulse supply.

*Cell culture*: Before culturing cells, the device was sterilized through immersing in 70 % alcohol for 30 min. The chip will be incubated by the 20 ug/mL FN for 2 h to improve the cell adhesion, then remove the extra solution. PC-3 cells (CRL-1435) were purchased from the American Type Culture Collection (ATCC) and maintained in RPMI-1640 medium (Gibco) supplemented with 10% fetal bovine serum and standard antibiotics (penicillin 100 U/ml and streptomycin 100 µg/ml). Cells were cultured at 37°C in a humidified atmosphere containing 5% $CO_2$ and were routinely tested for mycoplasma contamination. Cells were passaged at a ratio of 1:10–1:20 every 4–5 days using 0.05% Trypsin-EDTA with phenol red (Gibco). The cell sample for SEM imaging was prepared with 2.5% glutaraldehyde fixation and epon resin polymerization. Negative PBS solution was applied as liquid matrix on Raman measuring live cells.

*Raman measurement*: Raman measurements were obtained by a Renishaw inVia Raman spectrometer with Leica 63×/0.9 NA water immersion objective. And 785 nm laser power is approximately 9.5 mW with an exposure time of 0.5 s. Waterfall images were generated based on Raman spectra was proceeded through cosmic-ray-removal, subtract baseline in advance.

*Fluorescence imaging and intensity normalization*: Fluorescent images were captured via Zeiss Axio Scope.A1 upright fluorescence microscope and the data was processed by ZEISS ZEN 3.5 (blue edition). The PI fluorescence intensity level in Figure 4 was normalized according to equation as below,

$$I_{Norm} = \frac{I_t - I_B}{I_{Max} - I_B} \times 100\%$$

where $I_t$ and $I_B$ are the average PI fluorescent intensity of the cell at time $t$ and that of the background (before applying voltage), respectively. $I_{Max}$ represents for the maximum average intensity achieved for the cells that were killed under high amplitude pulse, which means cell membrane has been disrupted, and the amount of diffused PI molecules achieves maximum volume.

Acknowledgments

The research leading to these results received funding from University of Oulu, digital Raman assay for single-molecule protein sequencing in single live cells (LiveProSeq) project and Jane and Aatos Erkko Foundation, next generation molecular sensing (NGSens) project.

Conflict of interest

The authors declare no conflict of interest.